\documentclass[a4paper,11pt]{article}
\usepackage{jcappub}
\usepackage{graphicx}

\subheader{Accepted by JCAP}

\title{Cold or Warm? Constraining Dark Matter with Primeval Galaxies and Cosmic Reionization after \textsl{Planck}}
\author[a,b,c,d]{A. Lapi}
\author[a,c,d]{L. Danese}
\affiliation[a]{SISSA, Via Bonomea 265, 34136 Trieste, Italy}
\affiliation[b]{Dip. Fisica, Univ. `Tor Vergata', Via Ricerca Scientifica 1, 00133 Roma, Italy}
\affiliation[c]{INAF-Osservatorio Astronomico di Trieste, Via Tiepolo 11, 34131 Trieste, Italy}
\affiliation[d]{INFN-Sezione di Trieste, Via Valerio 2, 34127 Trieste, Italy}
\emailAdd{lapi@sissa.it}
\emailAdd{danese@sissa.it}

\abstract{Dark matter constitutes the great majority of the matter content in
the Universe, but its microscopic nature remains an intriguing mystery, with
profound implications for particle physics, astrophysics and cosmology. Here
we shed light on the longstanding issue of whether the dark matter is warm or
cold by combining the measurements of the galaxy luminosity functions out to
high redshifts $z\sim 10$ from the \textsl{Hubble Space Telescope}
with the recent cosmological data on the reionization history of the Universe
from the \textsl{Planck} mission. We derive robust and tight bounds on the
mass of warm dark matter particle, finding that the current data require it
to be in the narrow range between 2 and 3 keV. In addition, we show that a mass
not exceeding 3 keV is also concurrently indicated by astrophysical constraints
related to the local number of satellites in Milky Way-sized galaxies,
though it is in marginal tension with
analysis of the Lyman$-\alpha$ forest. For warm dark matter masses above
$3$ keV as well as for cold dark matter, to satisfy the \textsl{Planck} constraints
on the optical depth and not to run into the satellite problem would
require invoking astrophysical processes that inhibit galaxy formation in
halos with mass $M_{\rm H} \lesssim$ few $\times 10^8\,M_\odot$, corresponding to
a limiting UV magnitude $M_{\rm UV}\approx -11$.
Anyway, we predict a downturn of the galaxy luminosity function
at $z\sim 8$ faintward of $M_{\rm UV}\approx -12$, and stress that its
detailed shape is extremely informative both on particle physics and on
the astrophysics of galaxy formation in small halos. These expectations will
be tested via the \textsl{Hubble Frontier Fields} and with the advent of the
\textsl{James Webb Space Telescope}, which will enable probing the very faint
end of the galaxy luminosity function out to $z \sim 8-10$.}

\keywords{dark matter theory --- reionization --- particle physics - cosmology connection}

\begin{document}
\maketitle
\flushbottom

\section{Introduction}

Several astrophysical and cosmological probes have firmly established that
baryons --- which stars, planets, and (known) living creatures are made of
--- constitute only some $15\%$ of the total matter content in the Universe
\cite{Planck15}. The rest is in the form of `dark matter' (DM), which
basically does not interact with the baryons except via long-range
gravitational forces.

Still, no firm `direct' identification of the DM particles has been made so
far, although the progress in the sensitivity of detection experiments has
been increasing dramatically over the past decade, and there are claims of
intriguing, but not yet convincing, signals \cite{Bertone04,Feng10}. Neither
has robust evidence of DM been revealed from `indirect' searches in the sky,
by looking for $\gamma$-ray signals due to the annihilation or decay of DM
particles at the Galactic center, in nearby galaxies, and in the diffuse
$\gamma$-ray background \cite{Ackermann12,Ackermann14,Fermi15}. The detection
strategy that appears most promising today is the search for new physics in
accelerators, and in particular at the \textsl{Large Hadron Collider}
\cite{Mitsou15}.

Thus the microscopic nature of the DM largely remains a mystery.
Theoretically, based on extensions of the Standard Model, several candidate
particles with different masses and properties have been proposed
\cite{Bertone10}; relevant examples are axions, sterile neutrinos, and weakly
interacting massive particles (WIMPs) like the neutralino (the lightest
supersymmetric particle). In particular, WIMPs with masses in the GeV range
have long been considered a paradigm at the heart of the standard
cosmological model. Such massive DM particles are `cold' (hence the acronym
CDM), i.e., non-relativistic at the epoch of their decoupling from the
interacting particles in the Universe, and feature negligible free-streaming
velocities from the initial perturbations of the cosmic density field.

CDM leads to a bottom-up structure formation, according to which bound DM
structures called `halos' grow hierarchically in mass and sequentially in
time, with small clumps forming first and then stochastically merging
together into larger and more massive objects \cite{Frenk12}. These halos
provide the gravitational potential wells where baryonic matter can settle in
equilibrium, and via a number of complex astrophysical processes (e.g.,
cooling, star formation, energy feedback, etc.) originate the luminous
structures that populate the visible Universe.

On large cosmological scales, data on galaxy clustering and on the cosmic
microwave background confirm the above picture and are consistent with the
CDM paradigm; recently, this has been tested to an unprecedented accuracy by
the \textsl{Planck} mission \cite{Planck15}. However, on small scales the CDM
paradigm has been challenged by at least two longstanding issues: the
cusp-core controversy, and the missing satellite problem. The former arises
because N-body simulations predict the inner density profile of DM halos to
be cuspy \cite{Navarro97}, whereas observations find them cored
\cite{Salucci12}, i.e., round. The second problem comes because N-body
simulations predict, for Milky Way-sized halos, the existence of a number of
subhalos substantially larger than that of satellites found in our Galaxy
\cite{Boylan12}.

On the one hand, it has been suggested that such issues can be alleviated
within the standard CDM framework by accounting for the interplay between DM
and baryons during the galaxy formation process
\cite{ElZant01,Tonini06,Pontzen14}. On the other hand, they may also be
solved by considering a different microscopic scenario for the DM particle;
one appealing possibility is Warm Dark Matter (WDM) with masses in keV range,
as provided by, e.g., a sterile neutrino. Being lighter than CDM, WDM
particles remain relativistic for longer in the early universe and retain an
appreciable residual velocity dispersion; so they more easily free-stream out
from small-scale perturbations, suppressing the formation of subhalos and
originating flat density distributions at the halo centers
\cite{Bode01,Lovell14}. The strength of these effects depends on the WDM
particle mass; the lighter the particle mass $M_{\rm WDM}$, the larger the
free streaming scale $\propto M_{\rm WDM}^{-1.15}$.

From an astrophysical perspective, the free streaming scales of the WDM
particles can be indirectly probed by observations of the Lyman-$\alpha$
absorption produced by small clumps of neutral hydrogen in the spectra of
distant quasars, the so called Lyman-$\alpha$ forest; lower bounds on the WDM
mass $M_{\rm WDM}\gtrsim 2-3$ keV have been derived \cite{Seljak06,Viel13}.
However, these analyses suffer from the uncertainties in the astrophysical
modeling and numerical rendering of the intergalactic medium and of the
intervening absorbers.

Here we exploit an alternative, direct probe of the WDM mass
\cite{Pacucci13,Schultz14}, that is provided by the statistics of primeval
galaxies observed out to $z \sim 10$, supplemented with the data on the
reionization history of the Universe from cosmological observations; we
recall that reionization is the process by which the intergalactic medium has
transitioned again to an ionized state (it was already fully ionized before
the epoch of recombination, when the Universe was younger than 380.000 yr)
due to the radiation from the first astrophysical sources.

In a nutshell, the argument runs as follows. Faint galaxies at $z \gtrsim 4$
are the main responsible for the cosmic reionization (although faint active
galactic nuclei may also contribute \cite{Giallongo15,Madau15}\footnote{Given
their number density much smaller than protogalaxies' (see also
\cite{Weigel15}), AGN-driven reionization requires an escape fraction $f_{\rm
esc}\approx 1$.}), whose history has been recently constrained by the
\textsl{Planck} mission \cite{Planck15} in terms of integrated optical depth
$\tau_{\rm es}\approx 0.066\pm 0.012$ for Thomson scattering to the cosmic
microwave background\footnote{The analysis of only \textsl{Planck} data (TT,
TE, EE spectra + low-$\ell$ polarization + lensing) yields $\tau_{\rm
es}\approx 0.063\pm 0.014$, while when external data (e.g., BAOs) are added,
the value quoted in the main text is found. We stress that in the
$\Lambda$CDM cosmology, $\tau_{\rm es}$ is appreciably degenerate with the
amplitude $A_s$ of the primordial perturbation spectrum, but the associated
variations are still within its $2\sigma$ uncertainty; finally, $\tau_{\rm
es}$ is essentially unaffected in minimal extensions of the standard
$\Lambda$CDM model by the effective neutrino number $N_{\rm eff}$, by
dynamical dark energy $w_\Lambda$, etc.}. This measurement gauges the level
of the ionizing background from primeval galaxies, and in turn (although with
some assumptions to be discussed next) their number density. These galaxies
reside in small halos, so that their number density can provide a direct test
of the WDM free-streaming scale, and sets bounds on the particle mass.

Our working plan is straightforward: in Section 2 we present the reionization
history of the Universe as inferred from high-redshift galaxy statistics and
cosmological data; in Section 3 we derive the related constraints on the WDM
particle mass; in Section 4 we discuss the connection between the microscopic
nature of DM and the macroscopic, astrophysical properties of galactic halos;
in Section 5 we summarize our findings.

Throughout the work we adopt the standard, flat cosmology \cite{Planck15}
with matter density parameter $\Omega_M = 0.31$, baryon density parameter
$\Omega_b = 0.05$, Hubble constant $H_0 = 100\, h$ km s$^{-1}$ Mpc$^{-1}$
with $h = 0.68$, and mass variance $\sigma_8=0.81$.

\section{Reionization history}

Our starting point is constituted by the UV luminosity function of galaxies
at redshift $z \gtrsim 4$, illustrated in Fig. 1. Circles refer to data
points from observations with the \textsl{Hubble Space Telescope}
\cite{Bouwens15} (see also \cite{McLeod15}), while the solid lines are their
analytic renditions in terms of continuous Schechter functions ${\rm d}N/{\rm
d}M_{\rm UV} \propto 10^{-0.4\,(M_{\rm UV}-M_{\rm
UV}^\star)\,(\alpha+1)}\times \exp[-10^{-0.4\,(M_{\rm UV}-M_{\rm
UV}^\star)}]$; these feature a faint-end powerlaw slope $\alpha$ steepening
from $-1.7$ to $-2.3$ as $z$ increases from $4$ to $10$, and an exponential
cutoff brightward of a characteristic magnitude $M_{\rm UV}^\star$
correspondingly ranging from $-21.5$ to $-20.5$. We recall that the UV
magnitude and luminosity are related by $M_{\rm UV}\approx 5.9 - 2.5\,\log
\nu\,L_{\rm UV} [L_\odot]$, where $\nu$ is the frequency corresponding to
$1550$ Angstroms, and $L_\odot \approx 3.8\times 10^{33}$ erg s$^{-1}$ is the
bolometric solar luminosity.

The dashed lines illustrate the outcomes after correction for dust extinction
estimated from the UV continuum slope \cite{Bouwens14}. Dust correction is
irrelevant for magnitudes fainter than $M_{\rm UV} \approx -17$, but it is
useful to provide an estimate of the intrinsic star formation rate (SFR) in
primeval galaxies. The relation between UV magnitude and SFR depends on the
initial mass function (IMF), i.e., on the distribution of stellar masses
formed per unit SFR. For the locally observed Chabrier \cite{Chabrier03} IMF,
the relation $M_{\rm UV} \approx 18.5-2.5\, \log {\rm SFR} [M_{\odot}$
yr$^{-1}]$ holds, and has been used in labeling the upper scale of Fig. 1.

Based on the observed luminosity functions (without dust-correction), we then
compute the reionization history of the Universe, in terms of the evolution
with redshift of the electron scattering optical depth $\tau_{\rm es}$. In
brief, this is performed as follows. Firstly, the ionization rate
\begin{equation}
\dot N_{\rm ion}\approx f_{\rm esc}\, k_{\rm ion} \, \int^{M_{\rm
UV}^{\rm lim}}{\rm d} M_{\rm UV}\, {{\rm d}N\over {\rm d}M_{\rm UV}} \, L_{\rm UV}~
\end{equation}
is found on multiplying the integral of the UV luminosity functions down to a
limiting UV magnitude $M_{\rm UV}^{\rm lim}$ by the number $k_{\rm
ion}\approx 2\times 10^{56}$ of ionizing photons yr$^{-1}$ $L_\odot^{-1}$ as
appropriate for a Chabrier IMF, and again by the average escape fraction
$f_{\rm esc}\approx 0.2$ of ionizing photons from the interstellar medium of
primeval galaxies \cite{Mao07,Dunlop13,Robertson15}. Then, the ionization
rate is inserted into the standard evolution equation of the HII ionizing
fraction
\begin{equation}
\dot Q_{\rm HII} = {\dot N_{\rm ion}\over \bar n_{\rm H}}-{Q_{\rm HII}\over
t_{\rm rec}}
\end{equation}
that takes into account the competition between ionization and recombination
processes \cite{Madau99,Ferrara15}. In the above equation $\bar n_{\rm
H}\approx 2\times 10^{-7}\, (\Omega_b h^2/0.022)$ cm$^{-3}$ is the mean
comoving hydrogen number density. In addition, the recombination timescale
reads $t_{\rm rec}\approx 3.2$ Gyr $[(1+z)/7]^{-3}\, C_{\rm HII}^{-1}$, where
the case B coefficient for an IGM temperature of $2\times 10^4$ K has been
used; this timescale crucially depends on the clumping factor of the ionized
hydrogen, for which a fiducial value $C_{\rm HII}\approx 3$ is adopted
\cite{Pawlik13}. We shall discuss later the dependence of our results on
these parameters. Finally, the electron scattering optical depth is obtained
by integrating the ionized fraction over redshift
\begin{equation}
\tau_{\rm es}(z) = c\, \sigma_{\rm T}\,\bar n_{\rm H}\int^z{\rm d}z'\,f_e\,Q_{\rm HII}(z')
(1+z')^2 \, H^{-1}(z')~;
\end{equation}
here $H(z)=H_0\,[\Omega_M\,(1+z)^3+1-\Omega_M]^{1/2}$ is the Hubble
parameter, $c$ is the speed of light, $\sigma_{\rm T}$ the Thomson cross
section and $f_e$ the number of free-electron (computed assuming double
Helium ionization at $z\lesssim 4$.).

The outcome is illustrated in Fig. 2 for three representative values of the
limiting UV magnitude at the faint end: red lines are for $M_{\rm UV}
\lesssim -17$, green lines refer to $M_{\rm UV} \lesssim -13$ and blue lines
to $M_{\rm UV} \lesssim -11$; these produce integrated optical depths
$\tau_{\rm es}$ covering the $1\sigma$ region measured by \textsl{Planck},
with the value $M_{\rm UV}^{\rm lim}\approx -13$ approximately yielding the
\textsl{Planck} best fit $\tau_{\rm es}\approx 0.066$ \cite{Bouwens15b} (to
be precise, we obtain asymptotically $\tau_{\rm es}\approx 0.06$, while we
find $\tau_{\rm es}\approx 0.055$ if truncating the cosmic SFR at $z\gtrsim
8-10$, cf. \cite{Robertson15}). For reference, the dotted line represents the
optical depth expected in a fully ionized Universe up to redshift $z$; this
is to show that the bulk of the reionization process occurred at $z \sim
8-10$ and was almost completed at $z \sim 6$ \cite{Schull12}. Note that from
this perspective, the detailed behavior of the luminosity functions at $z
\gtrsim 10$ (that have been computed by extrapolation of the lower-redshift
behavior), and the related ionizing background, are only marginally relevant.
This is also apparent from the inset of Fig. 2, where the evolution with
redshift of the ionizing fraction $Q_{\rm HII}$ is illustrated, and
confronted with various observational constraints \cite{Robertson15}.

We note that the range $M_{\rm UV} \lesssim -17$ roughly corresponds to the
luminosity function currently observed at $z \sim 4-8$ and already yields
$\tau_{\rm es}\approx 0.045$. Going fainter to reproduce the \textsl{Planck}
best-fit value or the $1\sigma$ upper bound requires instead an extrapolation
into magnitude ranges currently inaccessible to observations at these
redshifts. One may wonder whether these extrapolations down to fainter
luminosities are reasonable. Actually, at lower redshift $z \sim 2$ the faint
end of the luminosity function has been explored down to $M_{\rm UV} \approx
-13$ thanks to gravitational lensing by a foreground galaxy cluster
\cite{Alavi14}; the data, illustrated by the magenta stars in Fig. 1,
indicates that the faint portion of the luminosity function keeps rising,
similarly to what happens with our extrapolations at higher $z$.

\section{Constraints on WDM particle mass}

We now aim at confronting the luminosity function of primeval galaxies with
the number density of the host DM halos. In Fig. 3 we illustrate the mass
function of halos ${\rm d}N/{\rm d}\log M_{\rm H}$ at the relevant redshift
$z\sim 8$ where the bulk of the reionization process occurs. We compute this
quantity based on analytic expressions for the halo mass function from the
excursion set theory \cite{Sheth01,Lapi13}, that approximate very well the
outcomes of $N-$body simulations for CDM \cite{Tinker08}; for the WDM cases,
we compute the mass variance (that enters the excursion set mass function) by
adopting a sharp filter in $k-$space, calibrated as prescribed by
\cite{Benson13,Schneider13} to reproduce the outcomes of WDM simulations at
different redshifts \cite{Angulo13,Schneider13,Bose15}.

In Fig. 3 the solid black line refers to the CDM case, while the colored
lines show the outcomes for different WDM masses $M_{\rm WDM} = 1, 2, 3$ keV.
For WDM the halo mass function is depressed below the free streaming mass of
the particles, that amounts to about $10^8\,M_\odot$ for $M_{\rm WDM}\approx
2$ keV and roughly scales as $M_{\rm WDM}^{-3.5}$. Note that we quote here
the Fermi$-$Dirac mass, i.e., the mass that the WDM particles would have if
they were thermal relics (decoupled in thermal equilibrium); this is
convenient because the masses of WDM particles produced in different
microscopic scenarios can be easily related to this quantity
\cite{Kusenko09}.

We should mention that the behavior of the halo mass function below the
free-streaming mass is somewhat debated, with various simulations and
analytic approaches providing quite different results
\cite{Benson13,Schneider13,Bose15}, especially at high redshifts $z\gtrsim
3$. In particular, the effective pressure from the residual WDM velocity
dispersion can make the truncation at small masses more dramatic; the dotted
colored lines in Fig. 3 are empirical renditions of these more extreme
behaviors \cite{Smith11}.

Now we come to the key point. In Fig. 4 we present the integrated number
density of DM halos $N(>\log M_{\rm H})$ with mass larger than $M_{\rm H}$ at
the relevant redshift $z \sim 8$. This is to be compared with the integrated
number density of faint galaxies down to the three limits $M_{\rm UV}^{\rm
lim}\approx -17$, $-13$, $-11$ relevant for reionization (cf. Fig. 2), which
are illustrated by the labeled grey shaded areas. This comparison highlights
that for small WDM particle masses, the number density in halos simply does
not attain the levels implied by the galaxy luminosity functions.
Specifically, it is evident that the observed galaxy number density down to
$M_{\rm UV} \approx -17$ straightforwardly constrains the WDM particle masses
to be $M_{\rm WDM}\gtrsim 1$ keV; values lower than this limit are ruled out,
essentially in a model-independent way. In fact, comparison between the solid
and dotted lines highlights that the uncertainty in the behavior of the halo
mass function below the free-streaming mass scale does not affect our
conclusion.

Moreover, as discussed above, the reionization history of the Universe
implied by the recent \textsl{Planck} data for $\tau_{\rm es}$ requires the
luminosity function to be extrapolated at least up to $z \sim 10$ with a
steep slope $\alpha\approx -2$ down to $M_{\rm UV} \approx -13$,
strengthening the constraints on the WDM mass to $M_{\rm WDM}\gtrsim 2$ keV.
On the other hand, consistency with the \textsl{Planck} $1\sigma$ upper limit
on $\tau_{\rm es}$ requires the luminosity function not to rise by much
beyond $M_{\rm UV} \approx -11$, implying the upper bound $M_{\rm WDM}
\lesssim 3$ keV.

We stress that such constraints are direct and robust, since they only
require minimal assumptions. Specifically, in Fig. 5 we show how the UV
limiting magnitude and the associated number density of galaxies required to
match the \textsl{Planck} best fit value on $\tau_{\rm es}\approx 0.066$,
change with the relevant parameters, namely, the IMF (via $k_{\rm ion}$ in
Eq. 2.1), the escape fraction $f_{\rm esc}$ of ionizing photons, the slope
$\alpha$ of the faint-end luminosity function and the clumping factor $C_{\rm
HII}$ of the intergalactic medium. The fiducial values are highlighted as
cyan stars, while higher or lower values are marked by labeled circles.
Moving upward in the plot from the fiducial value implies fainter magnitudes
and higher galaxy number density, hence tighter constraints on the WDM mass,
and viceversa. Note that here the parameters are varied one by one while the
others are kept fixed. However, their degeneracy in producing a given
reionization history can be highlighted by the approximate expression $f_{\rm
esc}\, k_{\rm ion}\,C_{\rm HII}^{-0.3}\, \Gamma[\alpha+2;10^{-0.4\,(M_{\rm
UV}^{\rm lim}-M_{\rm UV}^\star)}]\approx$ const (cf. also Eq. 6 in
\cite{Bouwens15b}), where $\Gamma[a;z]\equiv \int_z^\infty{\rm
d}x~x^{a-1}\,e^{-x}$ is the incomplete $\Gamma-$function; the most important
dependencies are on $f_{\rm esc}$ and on the limiting magnitude $M_{\rm
UV}^{\rm lim}$, given the noticeable observational uncertainties on these
parameters.

In more detail, switching from a Chabrier to a Salpeter IMF \cite{Salpeter55}
implies fewer ionizing photons being produced for a given SFR (parameter
$k_{\rm ion}$), hence the limiting magnitude required to reproduce the
\textsl{Planck} data increases from $M_{\rm UV}\approx -12.5$ to $\approx
-11$, the corresponding galaxy number density rises from $N(>\log M_{\rm H})
\approx -0.5$ to $\approx +0.5$, and the lower bounds on the WDM mass
strengthen from $M_{\rm WDM} \gtrsim 2$ keV to $\gtrsim 3$ keV, cf. Fig. 4.
Conversely, switching from the observed Chabrier IMF to a hypothetical
top-heavy IMF \cite{Greggio11} enhances the number of ionizing photons, and
causes the limiting magnitude to change from $M_{\rm UV}\approx -12.5$ to
$\approx -15.5$, the number density to decrease from $N(>\log M_{\rm
H})\approx -0.5$ to $\approx -1.75$, and the lower bounds on the WDM mass to
weaken from $M_{\rm WDM}\gtrsim 2$ keV to $\gtrsim 1.25$ keV; however, note
that such a top-heavy IMF would imply a correspondingly stronger metal and
dust enrichment of the interstellar medium in primeval galaxies already at
$z\sim 8$, which is not expected for these faint UV sources and in turn would
dramatically reduce their ionization efficiency.

For the other parameters: the clumping factor of the intergalactic medium has
a small impact on our results, while the escape fraction and the faint-end
slope of the luminosity function are more critical. Specifically, increasing
the escape fraction from the fiducial value of $0.2$ to $0.4$ or steepening
the slope of the luminosity function from the fiducial $\alpha\approx -2$ to
$-2.25$ would shift the allowed range of the WDM mass to $1.5 \lesssim M_{\rm
WDM}/{\rm keV }\lesssim 2.5$ keV.

However, values of $f_{\rm esc}\approx 0.2$ are considered conservative upper
limits, actually attained only for faint galaxies at high redshift $z \gtrsim
3$; this is demonstrated both by refined estimates
\cite{Nestor13,Cooke14,Vanzella15} of the ionizing emissivity from galaxies
out to $z\sim 4$ and by observations in local analogs of high-$z$ faint
galaxies \cite{Borthakur14}. Note that some recent hydrodynamical simulations
\cite{Ma15,Paardekooper15} aimed at studying protogalaxies suggest a slight
increase of the escape fraction with decreasing halo mass, due to a
combination of supernova feedback efficiency and dense gas column density,
but still produce values limited to $f_{\rm esc}\lesssim 0.2$ in the range of
low mass halos $M_{\rm H}\lesssim 10^9\, M_\odot$ relevant for reionization
(see Section 4).

Concerning the slope of the luminosity function, the current data at $z\sim
8$, limited to $M_{\rm UV} \lesssim -17$, indicates values close to our
fiducial $\alpha\approx -2$. Similar if slightly shallower slopes have been
found at lower $z\sim 4-6$ in the same magnitude range (cf. Fig. 1);
moreover, values of $\alpha\approx -1.8$ are measured at $z\sim 2$ from the
ultra-faint data \cite{Alavi14} that probes the luminosity function down to
$M_{\rm UV}\approx -13$. A slope of $\alpha\approx -1.8$ instead of $-2$ at
$z\sim 8$ would imply a lower bound on the WDM mass of $M_{\rm WDM} \gtrsim
5$ keV. In the near future, a closer assessment of the faint-end slope will
become available via the \textsl{Hubble Frontier Fields} Program (see
\texttt{http://www.stsci.edu/hst/campaigns/frontier-fields/)}, and with the
advent of the James Webb Space Telescope \cite{Gardner06}, which will allow
probing the galaxy luminosity function at $z \sim 8-10$ down to magnitudes
$M_{\rm UV}\approx -13$, and potentially even fainter by taking advantage of
gravitational lensing effects \cite{Atek15}.

\section{The astrophysicist's view}

As a final step, we now turn to connecting the microscopic nature of the DM
with the macroscopic astrophysical properties of primeval galaxies. For doing
this, we aim at deriving an average statistical relationship between the UV
magnitude of a galaxy (or the SFR) and its host halo mass; this can be done
via the abundance matching technique \cite{Vale04,Shankar06,Moster13}, i.e.,
by associating galaxies and halos with the same integrated number density.
The outcome at the relevant redshift $z\sim 8$ is illustrated in Fig. 6, both
for CDM and three representative values of the WDM mass. Comparisons between
the solid and dashed/dotted lines highlight that the outcome is insensitive
to dust-corrections in the luminosity functions and/or to the behavior of the
WDM mass function below the free-streaming mass length.

We find a relationship SFR $\propto M_{\rm H}^{1.45}$ between the SFR and
host halo mass, which is remarkably similar to what has been derived in the
higher mass range $10^{11} \lesssim M_{\rm H}/M_\odot \lesssim 10^{12}$ at
lower $z\lesssim 6$ \cite{Aversa15}; the slope is close to that expected for
galaxies where the SFR is regulated by the balance between cooling and energy
feedback from supernova explosions or stellar winds \cite{Finlator11}. In WDM
scenarios, the power-law relationship somewhat flattens and then stops at
around the free-streaming mass scale, just because low mass halos are not
formed, while it would extend down to very small halos in a CDM Universe.

As shown by numerical simulations \cite{Boylan14}, this is actually a serious
issue for CDM, because a substantial number of such small halos would survive
down to the present time as bound satellites of Milky Way-sized galaxies,
which are not observed. This is another way of presenting the missing
satellite problem mentioned at the beginning of this article. Solving the
issue for CDM requires invoking astrophysical processes that must severely
limit or even suppress galaxy formation in halos with masses $M_{\rm
H}\lesssim$ few $\times 10^8\,M_\odot$ (cf. the yellow-shaded area in Fig.
6), corresponding to UV magnitudes fainter than $M_{\rm UV}\approx-11$ at
$z\sim 8$; remarkably, this is the same limit concurrently indicated by the
$1\sigma$ upper bound on $\tau_{\rm es}$ from the \textsl{Planck} data. We
stress that the limiting magnitude $M_{\rm UV}\approx-11$ imposed by the
satellite problem is independent on $f_{\rm esc}$; once this limit is
respected, then $f_{\rm esc}$ cannot be much different from $0.2$ to yield
the Planck value of $\tau_{\rm es}\approx 0.066$. Physical processes
suppressing galaxy formation may include an increase of supernova feedback
efficiency in such small halos, or radiative feedback from the diffuse UV
background, or more complex phenomena \cite{Efstathiou92,Sobacchi13,Cai14}.
On the other hand, the satellite issue is naturally solved in WDM with
$M_{\rm WDM}\lesssim 3$ keV since halos with $M_{\rm H} \lesssim$ few $\times
10^8\,M_\odot$ are not formed at all, and even cooling/star formation
processes may be less efficient due to the lack of substructures
\cite{Pontzen14}.

We remark that the combination of these astrophysical constraints on the
satellite problem with the cosmological data on reionization further restrict
the plausible range of parameter values investigated in Fig. 5. For example,
a very low value of the escape fraction $f_{\rm esc}\approx 0.05$ as
sometimes claimed in literature \cite{Ma15} would require the reionization
process to be triggered at $z\gtrsim 10$ by very faint galaxies with $M_{\rm
UV}\approx -8$. This in turn would imply the WDM mass to exceed $5$ keVs and
hence to be indistinguishable from CDM. On the other hand, such very faint
galaxies residing in halo masses $M_{\rm H}\lesssim10^8\,M_\odot$ would
largely exceed the observed number of local satellites.

We recall from Fig. 2 that the primeval galaxies contributing most to the
cosmic reionization have UV magnitudes $M_{\rm UV}\approx -13$; in turn, from
Fig. 6, these are seen to be hosted in halos with $M_{\rm H}\approx
10^9\,M_\odot$ and to feature typical SFR$\approx 10^{-2}\, M_\odot$
yr$^{-1}$. The reionization process started at $z\sim 10$ and nearly
completed at $z\sim 6$, corresponding to a relatively short time lapse of a
few $\times 10^8$ yr. Thus the typical stellar masses accumulated in the
primeval galaxies responsible for reionization are $M_\star\approx$ few
$\times 10^6\, M_\odot$. Remarkably, this is consistent with the
extrapolation down to $M_{\rm UV}\approx -13$ of the $M_\star$ vs. $M_{\rm
UV}$ relationship \cite{Duncan14} currently estimated at $z\sim 7$ for
$M_{\rm UV}\lesssim -18$. The corresponding stellar mass density at $z\sim 8$
amounts to $\approx 10^6\, M_\odot$ Mpc$^{-3}$, which turns out to be $1/300$
of the integrated local value at $z \sim 0$. We note that the relative narrow
redshift range of the reionization between $z\sim 10$ and $z\sim 6$ as
inferred from the \textsl{Planck} data will ease tomographic mapping of the
HI distribution via the redshifted 21 cm line with the \textsl{Square
Kilometer Array} \cite{Carilli15,Carucci15}.

In Fig. 7 we provide specific predictions on the very faint end of the UV
luminosity function at $z\sim 8$, obtained by combining the $M_{\rm UV}$ vs.
$M_{\rm H}$ relationships of Fig. 6 with the halo mass function of Fig. 3;
note that the faint-end slope of the luminosity function mirrors that of the
halo mass function \cite{Bouwens15}. We check that for magnitudes $M_{\rm
UV}\lesssim -17$ the input luminosity function based on \textsl{Hubble Space
Telescope} data by \cite{Bouwens15} is recovered; moreover, at the very faint
end currently precluded to observations, we can predict the behavior of the
luminosity function. Specifically, for WDM masses $M_{\rm WDM}$ in the
relevant range from $2$ to $3$ keVs, we expect an abrupt downturn of the
luminosity function at magnitudes $M_{\rm UV}$ in the range from $-13$ to
$-11$ basically due to the lack of substructures.

On the other hand, for pure CDM the luminosity function would continue to
rise steeply, infringing the \textsl{Planck} constraints on $\tau_{\rm es}$
if $M_{\rm UV}\gtrsim -11$. However, as we have discussed above, baryonic
processes must also intervene not to run into the satellite problem; these
will flatten the luminosity function for halo masses $M_{\rm H}\lesssim$ few
$\times 10^8\, M_\odot$, corresponding (cf. Fig. 6) to UV magnitudes $M_{\rm
UV}\approx -12$. In this case the flattening is expected to be more gentle
because of the variance associated to the underlying astrophysical processes,
as represented by the dashed line in Fig. 7.

Again we stress that the reionization history of the Universe (cf. Fig. 2)
inferred from the \textsl{Planck} data \cite{Planck15} on $\tau_{\rm es}$ and
the missing satellite problem (cf. Fig. 6) \emph{concurrently} indicate that
the luminosity function must downturn in the range of magnitudes $-13\lesssim
M_{\rm UV}\lesssim -11$. In all these respects, the location and the detailed
shape of the luminosity function near the downturn would be extremely
informative both on particle physics and on the astrophysics of galaxy
formation in small halos \cite{Weisz14}. Whatever is its origin, DM nature or
baryon astrophysics, the downturn of the luminosity function is expected to
occur not far from the magnitudes that has been already observed at $z\sim 2$
\cite{Alavi14} thanks to gravitational lensing effects. At $z\sim 8$ the
relevant magnitude range will be probed in the next future via the
\textsl{Hubble Frontier Fields} and with the advent of the \textsl{James Webb
Space Telescope}.

\section{Conclusions}

In summary, we have shown that accurate observations on the number density of
primeval galaxies (cf. Fig. 1), supplemented with cosmological data
constraining the history of cosmic reionization (cf. Fig. 2), have a strong
potential for unveiling the elusive nature of dark matter. Specifically, we
have obtained a robust lower bound $M_{\rm WDM}\gtrsim 1$ keV from the number
density of primeval galaxies currently observed down to $M_{\rm UV}\approx
-17$ at $z\lesssim 8$ (cf. Figs. 3 and 4). In addition, the \textsl{Planck}
measurements on the electron scattering optical depth $\tau_{\rm es}\approx
0.066$ (and constraints on the evolution of the ionized fraction from various
astrophysical probes) can be reproduced with standard, observationally
supported (at least for $z\lesssim 3$) assumptions on the escape fraction
$f_{\rm esc}\approx 0.2$, the clumping factor $C_{\rm HII}\approx 3$, and the
faint-end slope $\alpha\approx -2$ of the luminosity function extrapolated
down to the UV limiting magnitude $M_{\rm UV}\approx -13$.

Such values imply an even tighter constraint on the WDM mass $M_{\rm
WDM}\gtrsim 2$ keV; on the other hand, an upper bound $M_{\rm WDM}\lesssim 3$
keV is concurrently indicated by the $1\sigma$ upper limit on the
\textsl{Planck} data on $\tau_{\rm es}$ and by astrophysical constraints
related to the local number of satellites in Milky Way-sized galaxies (cf.
Fig. 6), though it is in marginal tension with analysis of the Lyman$-\alpha$
forest (see also discussion by \cite{Brooks14}). This shows that, in the
perspective of astrophysics and cosmology, only WDM endowed with particle
masses just around $2-3$ keV is a relevant alternative to the CDM case. On
the other hand, for WDM masses above $3$ keV as well as for CDM, to comply
with the \textsl{Planck} constraints on $\tau_{\rm es}$ and not to run into
the satellite problem would require invoking astrophysical processes that
inhibit galaxy formation in halos with mass $M_{\rm H} \lesssim$ few $\times
10^8\,M_\odot$, corresponding to $M_{\rm UV}\approx -11$.

As a specific prediction, we expect a downturn of the galaxy luminosity
function at $z\sim 8$ faintward of $M_{\rm UV}\approx -12$, which should be
more abrupt if it is due to the free-streaming scale associated with WDM, or
more progressive if due to astrophysical processes in a CDM Universe (cf.
Fig. 7). Such a downturn is expected to occur not far from the magnitudes
that has been already observed at $z\sim 2$ \cite{Alavi14} thanks to
gravitational lensing effects. These expectations will be tested via the
\textsl{Hubble Frontier Fields} and with the advent of the \textsl{James Webb
Space Telescope}, which will enable probing the very faint end of the galaxy
luminosity function out to $z \sim 8-10$. Specifically, its detailed shape
will eventually answer the question: which is the main cause regulating
galaxy formation within small galaxy halos, particle physics or astrophysics?

\acknowledgments This work has been supported in part by the MIUR PRIN
2010/2011 `The dark Universe and the cosmic evolution of baryons: from
current surveys to Euclid', by the INAF PRIN 2012/2013 `Looking into the
dust-obscured phase of galaxy formation through cosmic zoom lenses in the
Herschel Astrophysical Terahertz Large Area Survey'. We acknowledge our
referee for helpful comments. We are grateful to R. Aversa, A. Cavaliere, C.
Baccigalupi, G. De Zotti, P. Salucci, and A. Schneider for discussions, and
to J. Miller for critical reading. A.L. thanks SISSA for warm hospitality.

\clearpage
\begin{figure*}
\centering
\includegraphics[width=14cm]{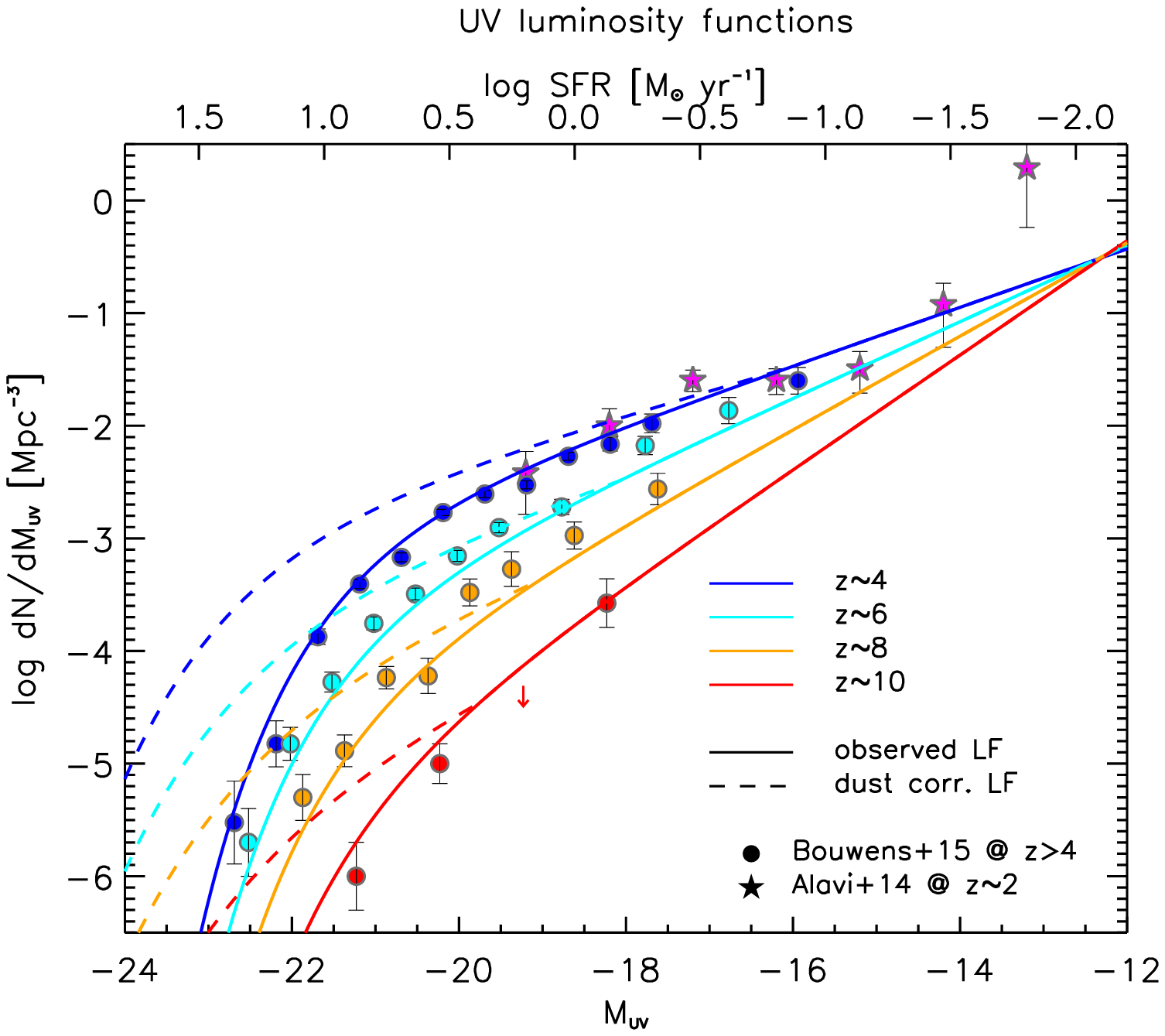}
\caption{The UV luminosity function. Circles refer to data points from the
\textsl{Hubble Space Telescope} \cite{Bouwens15} at $z \sim 4$ (blue), $6$
(cyan), $8$ (orange), and $10$ (red), with the solid lines illustrating their
analytic renditions in terms of Schechter functions; the dashed lines include
dust-correction estimated from the UV continuum spectral slope
\cite{Bouwens14}. The purple stars show the ultra-faint luminosity function
at $z \sim 2$ measured with the aid of gravitational lensing by a foreground
galaxy cluster \cite{Alavi14}. The upper scale refers to the SFR associated
with the dust-corrected UV magnitude.}
\end{figure*}

\clearpage
\begin{figure*}
\centering
\includegraphics[width=14cm]{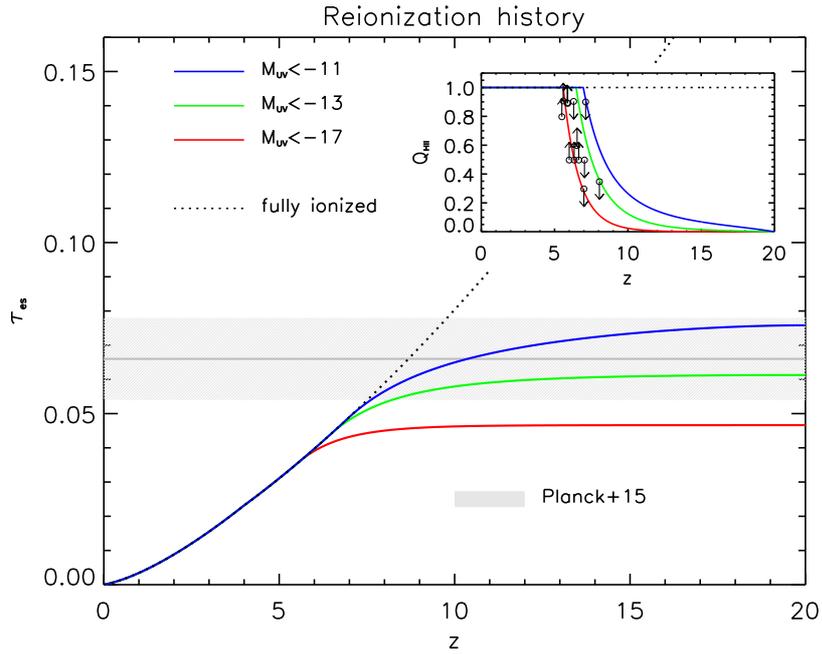}
\caption{The evolution with redshift of the electron scattering optical depth
$\tau_{\rm es}$. Solid lines illustrate the result by using the luminosity
functions down to a limiting magnitude $M_{\rm UV}= -17$ (red), $-13$
(green), and $-11$ (blue), see text. For reference, the black dotted line
refers to a fully ionized Universe up to redshift $z$. The grey shaded area
shows the measurement (with $1\sigma$ uncertainty region) from the
\textsl{Planck} Collaboration \cite{Planck15}. In the inset, the
corresponding evolution of the ionized fraction $Q_{\rm HII}$ is plotted,
together with upper and lower limits from various observations
\cite{Robertson15}.}
\end{figure*}

\clearpage
\begin{figure*}
\centering
\includegraphics[width=14cm]{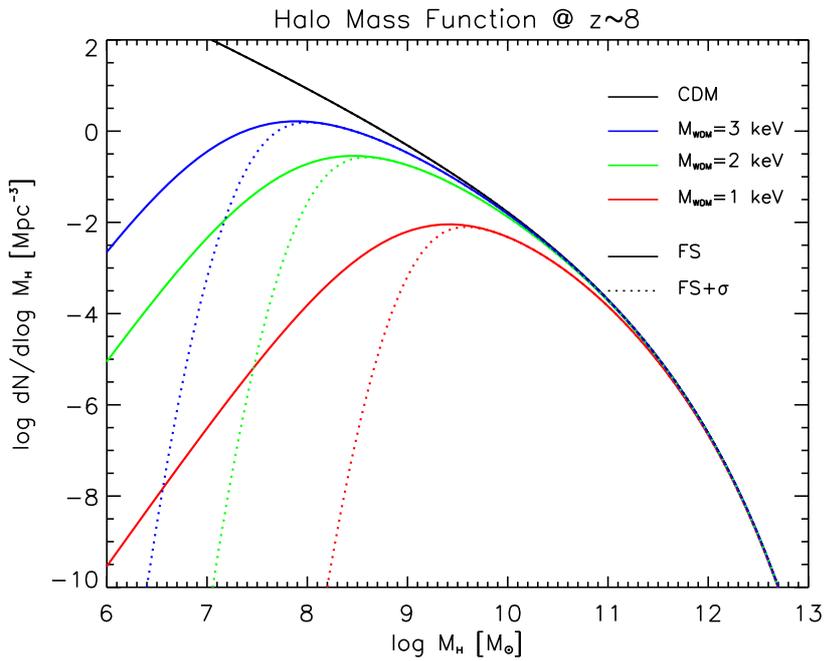}
\caption{The halo mass function at $z\sim 8$, for different values of the WDM
mass $M_{\rm WDM} = 1$ (red), $2$ (green), and $3$ (blue) keV. The solid
lines include only the free-streaming effect, while dotted lines also take
into account the effective pressure from residual velocity dispersion of the
WDM particles \cite{Bode01,Schneider13}. For reference, the black solid line
shows the halo mass function for standard CDM.}
\end{figure*}

\clearpage
\begin{figure*}
\centering
\includegraphics[width=14cm]{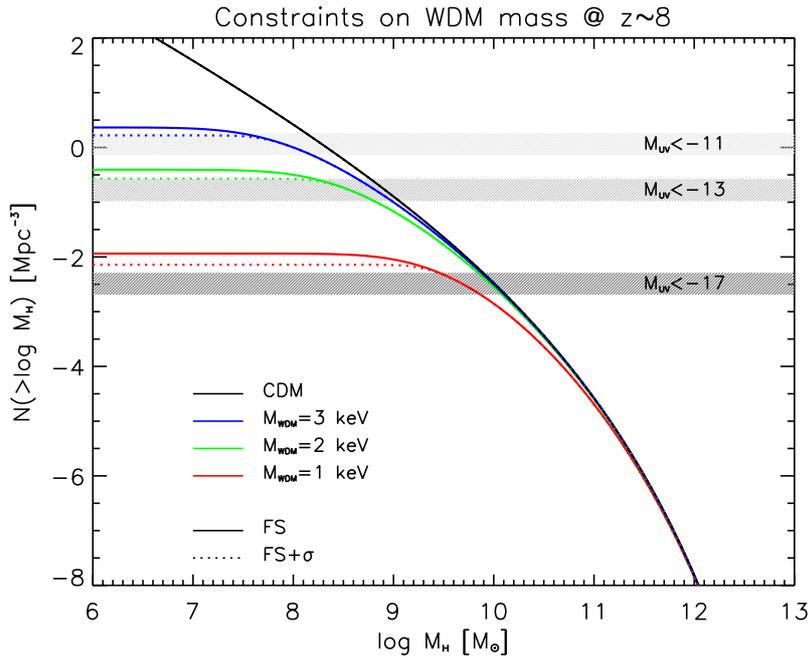}
\caption{Constraints on the WDM mass at $z \sim 8$ in terms of the comparison
between the integrated number density of halos and galaxies. The grey shaded
areas represent the integrated number density of galaxies down to limiting
magnitudes $M_{\rm UV}= -17$, $-13$, and $-11$, as labeled. The lines refer
to the integrated number density of halos, for different values of the WDM
mass $M_{\rm WDM}= 1$ (red), $2$ (green), and $3$ (blue) keV. The solid lines
include only the free-streaming effect, while the dotted lines also take into
account the effective pressure from the residual velocity dispersion of the
WDM particles. For reference, the black solid line shows the outcome for
standard CDM.}
\end{figure*}

\clearpage
\begin{figure*}
\centering
\includegraphics[width=14cm]{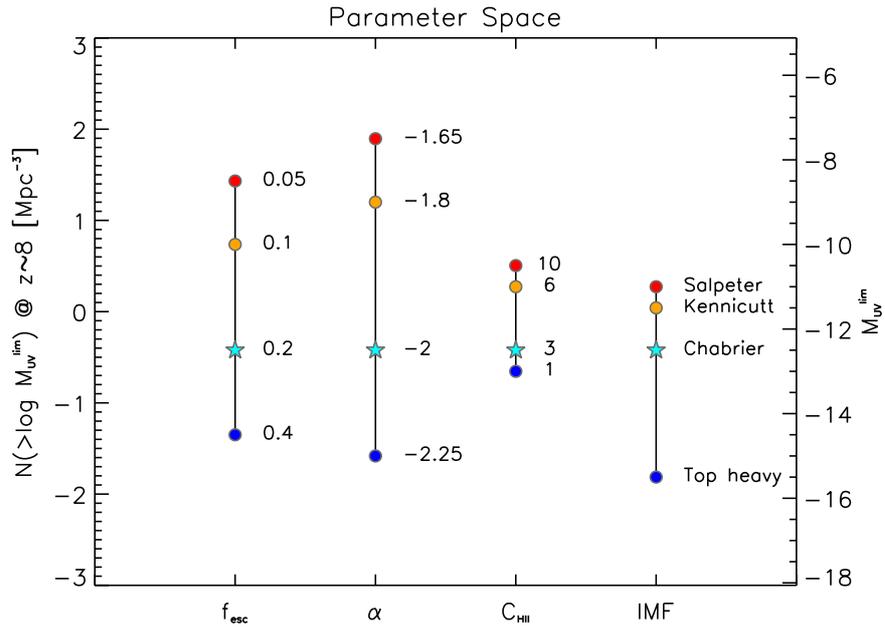}
\caption{Effects on our results of varying the basic assumptions concerning:
the escape fraction of ionizing photons $f_{\rm esc}$, the faint-end slope of
the UV luminosity function $\alpha$, the clumping factor $C_{\rm HII}$ of the
intergalactic medium, and the IMF. The right axis shows the limiting
magnitude, and the left axis shows the corresponding integrated galaxy number
density, required to match the \textsl{Planck} best fit of $\tau_{\rm
es}\approx 0.066$. Our fiducial values of the parameters are illustrated as
cyan stars, while the dots represent larger and lower values as labeled.}
\end{figure*}

\clearpage
\begin{figure*}
\centering
\includegraphics[width=14cm]{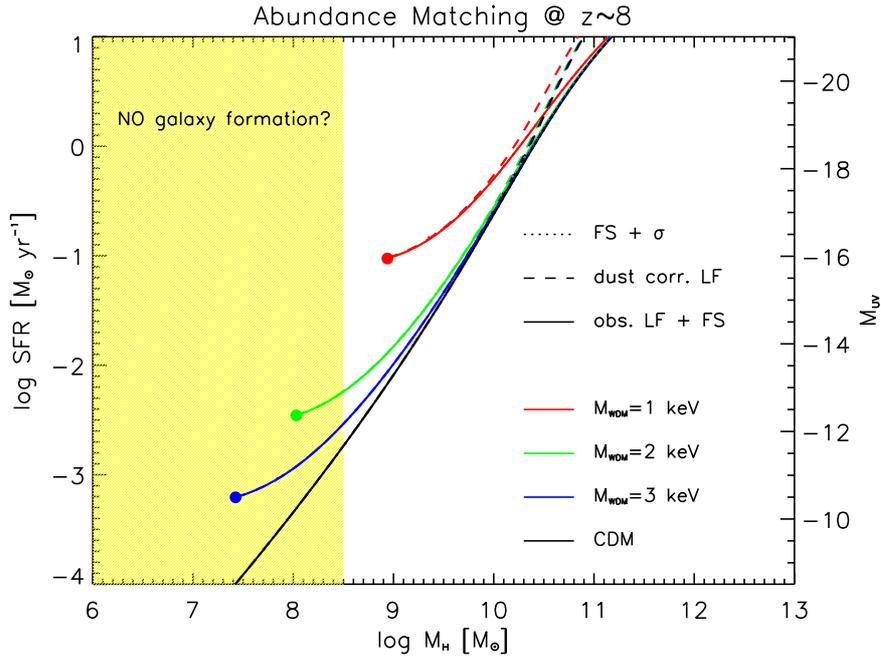}
\caption{The UV magnitude (right axis) and the SFR (left axis) plotted
against the halo mass at $z \sim 8$, obtained from the abundance matching
technique, for different values of the WDM mass $M_{\rm WDM} = 1$ (red), $2$
(green), and $3$ (blue) keV. The solid lines refer to the results obtained by
using the observed galaxy luminosity functions and the halo mass functions
with only free streaming effects included. The dashed lines were obtained
using the dust-corrected galaxy luminosity functions, while the dotted lines
(which turn out to be nearly coincident with the solid lines) were obtained
from the halo mass functions with inclusion of both free-streaming and
velocity dispersion effects. For reference, the black solid line shows the
outcome for standard CDM. In WDM scenarios, the colored curves are
interrupted at the free-streaming mass scale (indicated by a filled dot). The
yellow shaded area marks the region
 where galaxy formation must be inefficient so as not to produce,
 according to observations and simulations \cite{Boylan14}, too many satellites
 in Milky Way-sized halos at $z\sim 0$.}
\end{figure*}

\clearpage
\begin{figure*}
\centering
\includegraphics[width=14cm]{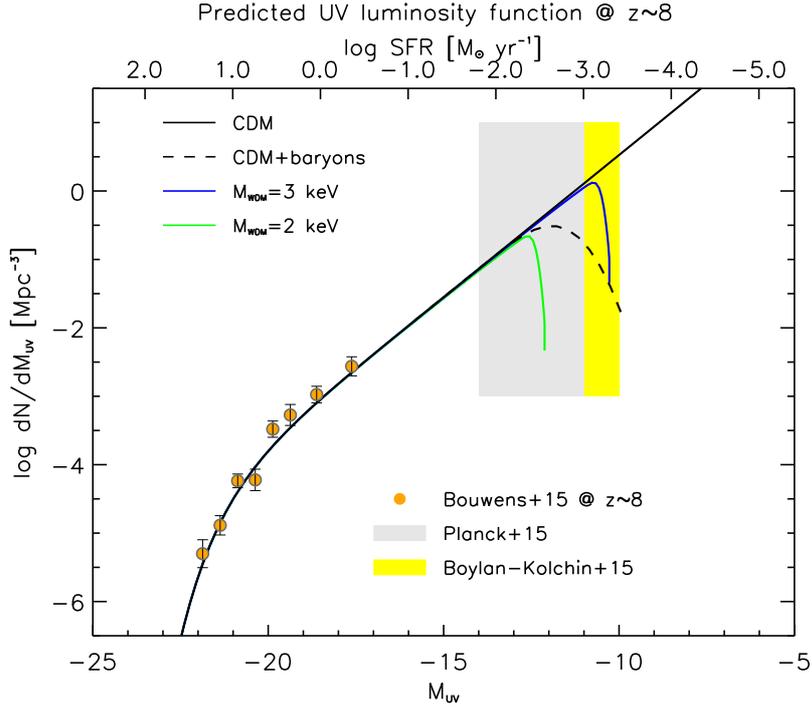}
\caption{Predictions on the very faint end of the UV luminosity function at
$z\sim 8$, obtained by combining the $M_{\rm UV}$ vs. $M_{\rm H}$
relationships of Fig.~6 with the halo mass function. Colored lines refer to
different values of the WDM mass $M_{\rm WDM} = 2$ (green) and $3$ (blue)
keV; solid black line is for pure CDM, while dashed black line includes
baryonic effects (see text) required not to run into the missing satellite
problem. Orange circles show the data at $z\sim 8$ \cite{Bouwens15}. The
shaded areas illustrate the range of magnitudes where the luminosity function
must downturn to be consistent with the reionization history of the Universe
from \textsl{Planck} \cite{Planck15} data (grey shade; cf. Fig.~2) and not to
run into the satellite problem (yellow shade; cf. Fig. 6).}
\end{figure*}

\end{document}